\documentclass[iop]{emulateapj}
\usepackage{amssymb,multirow,color,natbib}

\begin{document}
\title{Comment on the numerical measurements of the magnetohydrodynamic turbulence spectrum by A. Beresnyak (Phys. Rev. Lett. 106 (2011) 075001; MNRAS 422 (2012) 3495; ApJ 784 (2014) L20)} 
\author{Jean Carlos Perez$^1$}
\author{Joanne Mason$^2$}
\author{Stanislav Boldyrev$^{3,4}$}
\author{Fausto Cattaneo$^5$}
\affiliation{$^1$Space Science Center, University of New Hampshire,
  Durham, NH, 03824; jeanc.perez@unh.edu\\
  $^2$College of Engineering, Mathematics and
  Physical Sciences, University of Exeter, EX4 4QF, UK; j.mason@exeter.ac.uk\\
  $^3$Department
  of Physics, University of Wisconsin at Madison, 1150 University Ave,
  Madison, WI, 53706;  boldyrev@wisc.edu\\
  $^4$Kavli Institute for Theoretical Physics,
  University of California at Santa Barbara, Santa Barbara, CA, 93106\\
  $^5$Department of Astronomy and Astrophysics, University of Chicago,
  5640 S. Ellis Ave, Chicago, IL, 60637; cattaneo@flash.uchicago.edu }  

\date{\today}

\begin{abstract}
The inertial-interval energy spectrum of strong magnetohydrodynamic (MHD) turbulence with a uniform background magnetic field was observed  numerically to be close to $k^{-3/2}$ by a number of independent groups. A dissenting opinion has been voiced by \citet[][]{beresnyak_11,beresnyak_12,beresnyak_14} that the spectral scaling is close to~$k^{-5/3}$. The conclusions of these papers are however incorrect as they are based on numerical simulations that are drastically  unresolved, so that the discrete numerical scheme does not approximate the physical solution at the scales where the measurements are performed. These results have been rebutted in our more detailed papers \citep{perez_etal2012,perez_etal2014}; here, by popular demand, we present a brief and simple explanation of our major criticism of Beresnyak's work.  
\end{abstract}

\maketitle

The field-perpendicular inertial-interval spectrum of large-scale driven, homogeneous, strong incompressible MHD turbulence with a strong uniform magnetic field was observed  numerically to be close to~$-3/2$ in a number of studies \citep{maron_g01,haugen_04,muller_g05,mininni_p07,chen_11,mason_cb06,mason_cb08,perez_b10_2,perez_etal2012,chandran_14}. A dissenting opinion has been voiced by~\citet[][]{beresnyak_11} and repeated in \citep[][]{beresnyak_12,beresnyak_14} that the spectrum is actually close to~$-5/3$.


\begin{figure}[tbp]
\includegraphics[width=\columnwidth]{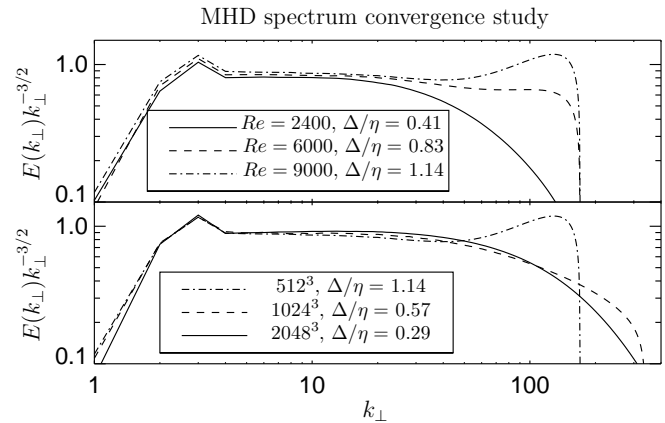}
\caption{\label{fig:spectra1}Convergence study of the numerical spectrum in MHD turbulence (an RMHD code is used; the numerical procedure is described in detail in \cite[][]{perez_etal2012}). Top: varying $Re$ at \emph{fixed} resolution $512^3$. The solid line represents the energy spectrum where the inertial interval is well resolved. The dash and dash-dotted lines represent similar simulations where the Reynolds number is increased so that the simulations become unresolved. As a result, at $k\gtrsim 15$ the 
 numerical solution does not approximate the physical one: the numerical spectrum steepens and then flattens closer to the cut-off scale, which is a purely numerical effect. Bottom: varying resolution at \emph{fixed} $Re=9000$. The dash-dotted line shows the spectrum of the most unresolved simulation 
where an unphysical distortion is present at small scales ($k\gtrsim 15$). This numerical distortion progressively disappears as the resolution is increased 
without changing the physical parameters of the simulations (the dashed-dotted and solid curves); the inertial interval now  extends to about $k\sim 30$.}
\end{figure}

An inspection of Beresnyak's work however demonstrates that its claims are based on drastically unresolved numerical simulations, which do not represent the physical solution. We illustrate this in  Fig.~\ref{fig:spectra1}. The solid line in the top panel represents the energy spectrum of a well resolved simulation in a $512^3$ mesh and Reynolds number $Re=2400$. The dash and dash-dotted lines show the same set up with larger Reynolds numbers $Re=6000$ and $9000$. 
In the latter two cases the scales at $k\gtrsim 15$ are significantly unresolved and affected by the proximity to the dealiasing cut-off $k_c={2\pi}/(3\Delta)$, where $\Delta$ is the grid size. 

The proximity to the $k$-space cutoff is known to distort the spectral behavior at small scales in hydrodynamic simulations  \cite[][]{cichowlas_2005,frisch_etal2008,connaughton2009,grappin2010}; the unresolved curves in 
Fig.~\ref{fig:spectra1} bear close similarity with those results as the ratio $\Delta/\eta$ gets closer to $1$, where $\eta$ is the dissipation scale \cite[$\eta$ and $Re$ are defined in e.g.,][]{perez_etal2012}.  
Such a numerical spectral distortion has been confused by  \citet[][]{beresnyak_11,beresnyak_12,beresnyak_14} with the inertial interval. A standard convergence test shown in the bottom panel of Fig.~\ref{fig:spectra1} illustrates 
that the distortion 
is eliminated by progressively increasing the numerical resolution to $1024^3$ and $2048^3$ keeping all the physical parameters unchanged.

\begin{figure}[tbp]
\includegraphics[width=\columnwidth]{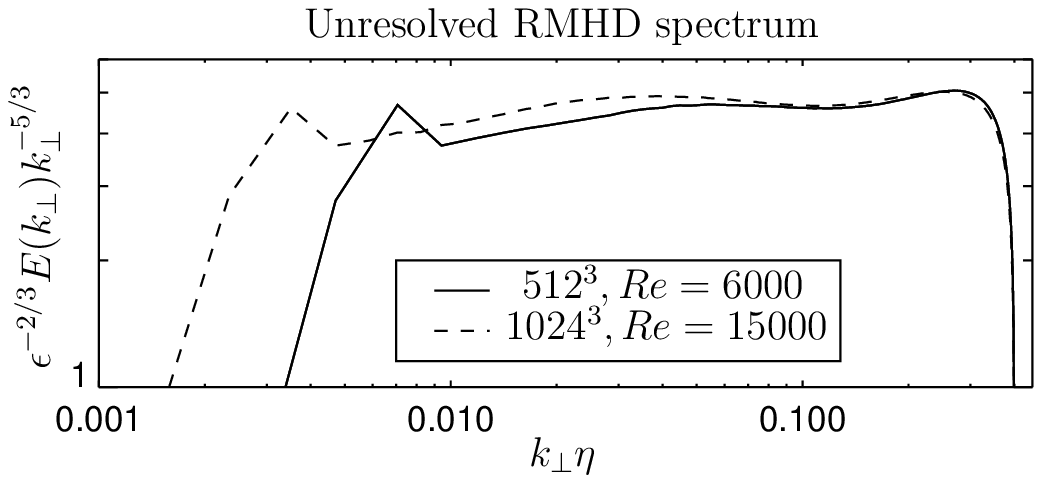}
\caption{\label{fig:spectra3}The scaling of
  the unresolved numerical spectrum in two runs with different resolutions and Reynolds numbers.  
The $Re$ numbers are chosen such that 
$\Delta/\eta \approx 0.83$, and this ratio is the same in both
  runs. Both simulations are essentially unresolved  at small scales, $k\eta\gtrsim 0.1$, where the solution of the numerical scheme scales differently from the inertial interval.} 
\end{figure}

In the resolved runs, the energy spectrum agrees with the $-3/2$ scaling; see the solid lines in Fig.~\ref{fig:spectra1}. However, in the unresolved runs, the scaling of the unresolved part of the spectrum 
is steeper and closer to $-5/3$, see Fig.~\ref{fig:spectra3}. This is not surprising: the scaling of the discrete numerical scheme does {\em not} have to agree with the scaling of the physical solution if the former does not approximate the latter \cite[e.g.,][]{perez_etal2012,perez_etal2014}. 
The scaling of the unresolved part in Fig.~\ref{fig:spectra3} 
was incorrectly attributed by  \citet[][]{beresnyak_11,beresnyak_12,beresnyak_14} to the scaling of the physical solution, as
the standard procedure (described in Fig.~\ref{fig:spectra1}) to test the numerical convergence  
was {\em not} followed in these works.

Due to the presence of the Alfv\'enic velocity scale, the Kolmogorov first self-similarity hypothesis cannot be formulated in MHD turbulence, and the energy spectrum cannot be established based solely on dimensional arguments (more discussion can be found in \cite{perez_etal2014}). In this case, numerical studies should be conducted with the utmost care. The incorrect conclusions of \cite[][]{beresnyak_11,beresnyak_12,beresnyak_14} could be avoided if their numerical simulations followed the standard procedures (as in Fig.~\ref{fig:spectra1}), where first and foremost, the numerical convergence is checked before any meaningful measurements are made.

\begin{acknowledgments}
This work was supported by NSF/DOE grant AGS-1003451, grant NNX11AJ37G from NASA's Heliophysics Theory Program,  DoE grant DE-SC0003888, NSF grant NSF PHY11-25915, and the NSF Center for Magnetic Self-Organization in Laboratory and Astrophysical Plasmas. High Performance Computing resources were provided by the Texas Advanced Computing Center (TACC) at the University of Texas at Austin under the NSF-XSEDE Project TG-PHY110016.
\end{acknowledgments}


\end{document}